\documentclass[aps,pra,preprint,groupedaddress,showpacs,floatfix]{revtex4-1}

\usepackage{graphicx}
\usepackage[english]{babel}

\usepackage{bm}
\usepackage{slashed}
\usepackage{fancybox}
\usepackage{amssymb}
\usepackage{mathrsfs,amsthm,amsmath}

\usepackage{epsfig}
\usepackage{graphicx}
\usepackage{color}

  \usepackage{slashed}
  \usepackage{url}
\usepackage{bm}
\usepackage{verbatim}
\usepackage{float}

\usepackage{stackrel}

\begin{document}

\newtheorem{theorem}{Theorem}
\newtheorem{corollary}{Corollary}
\def\endprf{\hfill  {\vrule height6pt width6pt depth0pt}\medskip}
\newcommand{\sgn}{\operatorname{sgn}}

\newtheorem{theo}{Theorem}

\title{Noncommutative geometry and stochastic processes}

\author{Marco Frasca}
\email[]{marcofrasca@mclink.it}
\affiliation{Via Erasmo Gattamelata, 3 \\ 00176 Roma (Italy)}

\begin{abstract}
The recent analysis on noncommutative geometry, showing quantization of the volume for the Riemannian manifold entering the geometry, can support a view of quantum mechanics as arising by a stochastic process on it. A class of stochastic processes can be devised, arising as fractional powers of an ordinary Wiener process, that reproduce in a proper way a stochastic process on a noncommutative geometry. These processes are characterized by producing complex values and so, the corresponding Fokker--Planck equation resembles the Schr\"odinger equation. Indeed, by a direct numerical check, one can recover the kernel of the Schr\"odinger equation starting by an ordinary Brownian motion. This class of stochastic processes needs a Clifford algebra to exist. In four dimensions, the full set of Dirac matrices is needed and the corresponding stochastic process in a noncommutative geometry is easily recovered as is the Dirac equation in the Klein--Gordon form being it the Fokker--Planck equation of the process.
\end{abstract}

\maketitle

\section{Introduction}

A comprehension of the link between stochastic processes and quantum mechanics can provide a better understanding of the role of space--time at a quantum gravity level. Indeed, noncommutative geometry, in the way Connes, Chamseddine and Mukhanov provided recently \cite{Chamseddine:2014nxa,Chamseddine:2014uma}, seems to fit well the view that a quantized volume yields a link at a deeper level of the connection between stochastic processes and quantum mechanics. This is an important motivation as we could start from a reformulation of quantum mechanics to support or drop proposals to understand quantum gravity and the fabric of space-time.

A deep connection exists between Brownian motion and binomial coefficients. This can be established by recovering the kernel of the heat equation from the binomial distribution for a random walk (Pascal--Tartaglia triangle) and applying the theorem of central limit \cite{weiss}. When an even smaller step in the random walk is taken a Wiener process is finally approached.  So, it is a natural question to ask what would be the analogous of Pascal--Tartaglia triangle in quantum mechanics\cite{fari}. This arises naturally by noting the apparent formal similarity between the heat equation and the Schr\"odinger equation. But this formal analogy is somewhat difficult to understand due to the factor $i$ entering into the Schr\"odinger equation. An answer to this question hinges on a deep problem not answered yet: Is there a connection between quantum mechanics and stochastic processes?  The formal similarity has prompted attempts to answer as in the pioneering work of Edward Nelson \cite{nels} and in the subsequent deep analysis by Francesco Guerra and his group \cite{Guerra:1981ie}. They dubbed this reformulation of quantum mechanics as ``stochastic mechanics''. This approach matches directly a Wiener process to the Schr\"odinger equation passing through a Bohm-like set of hydrodynamic equations and so, it recovers all the drawbacks of Bohm formulation. This view met severe criticisms motivating some researchers to a substantial claim that ``no classical stochastic process underlies quantum mechanics'' \cite{hang} showing contradiction with predictions of quantum mechanics. Subsequent attempts to partially or fully recover this view were proposed with non-Markovian processes \cite{skor} or repeated measurements \cite{bla1,bla2,bla3}.

In this paper we will show that a new set of stochastic processes can be devised that can elucidate such a connection \cite{fari,fra1}. We show their existence \cite{fra2} and we will determine how spin is needed also in the non-relativistic limit. Dirac equation for a free particle is also obtained. These processes are characterized by the presence of a Bernoulli process yielding the values $1$ and $i$, exactly as expected in the volume quantization in noncommutative geometry. In this latter case, it appears that a stochastic process on a quantized manifold is well represented by a fractional power of an ordinary Wiener process when this is properly defined through a technique at discrete time. For our aims we will use the simplest one: The Euler--Maruyama technique. A numerical test will yield the proof of existence for this class of stochastic processes. Also, the kernel of the Schr\"odinger equation is numerically obtained through an ordinary Brownian motion.


The paper is so strctured. In Sec.~\ref{sec2_0} we discuss noncommutative geometry in informal way, providing a general formula for a stochastic process on a quantized Riemannian manifold. In Sec.~\ref{sec2}, we introduce the fractional powers of a Wiener process and we solve the corresponding stochastic equation recovering the Wiener process we started from after squaring its square root. In Sec.~\ref{sec3}, we derive the formula for the square root of a Wiener process expressing it through more elementary processes: This shows the need for a Clifford algebra and the Fokker--Planck equation is obtained for a free particle. In Sec.~\ref{sec4}, we show numerically how the kernel of the Schr\"odinger equation is recovered by an ordinary Brownian motion just with the extraction of its square root. In Sec.~\ref{sec5}, we derive the Fokker--Planck equation in presence of a potential and specialize to the case of a harmonic oscillator. In Sec.~\ref{sec6}, we show how to recover a stochastic process on a noncommutative geometry taking the square root of more Wiener processes and using the algebra of the Dirac matrices. In Sec.~\ref{sec7}, we recover the Dirac equation as the Fokker--Planck equation for a square root process. Finally, in Sec.~\ref{sec8} conclusions are presented.

\section{noncommutative geometry and quantization of volume\label{sec2_0}}

\subsection{Definition of a noncommutative geometry}

Common wisdom on geometry implies that one has to cope with points and minimum paths between them. Indeed, the idea of geometry can be extended without the central concept of points but rather functions and introducing a redefinition of the concept of distance beside the well-known one from a Riemannian geometry. This reformulation is due to Alain Connes \cite{Connes:1994yd}. Essentially, one introduces a triple composed by an algebra of functions $\cal A$ with an involution operator like complex conjugation, playing the role of coordinates, a Hilbert space $\mathbb{L}^2$, that we take the space of the square-summable spinors, and a Dirac operator $D=i(\gamma\cdot\partial+\omega_\mu)$, being $\omega_\mu$ a spin connection, representing momenta. The algebra of functions has support on a Riemann manifold. When we change the algebra of functions with a noncommutative algebra of operators acting on the given Hilbert space, in the same way one quantize a classical theory, one gets a noncommutative geometry. So, a geometry is identified by the triple ($\cal A$,$\mathbb{L}^2$,D). A function $f$ belonging to $\cal A$ should satisfy the Lipschitz condition on the Riemann manifold given by
\begin{equation}
    {\rm Lip}(f):\qquad |f(x)-f(y)|\le L\cdot d_R(x,y)
\end{equation}
provided the constant $L$ exists and
\begin{equation}
   d_R(x,y)=\stackrel[\gamma]{\ }{\rm inf}\left(\int_\gamma ds\right)
\end{equation}
is the usual (geodesic) distance on a Riemann manifold that coincides with the well-known variational principle of minimum distance between two points. This grants some regularity properties of the functions in $\cal A$ and their gradient that is bounded. In this way, we can introduce a new definition of distance dependent just on the algebra of functions $\cal A$ and the Dirac operator. This is given by
\begin{equation}
   d(x,y) = \stackrel[f]{\ }{\rm sup}(|f(x)-f(y)|: f\in{\cal A},\lVert Df\rVert<1)
\end{equation}
where the condition on the Dirac operator plays a crucial role. In this way, one recovers the ordinary Riemann distance between points \cite{Bimonte:1994ch}. Indeed, one has for a spinor $\psi\in\mathbb{L}^2$
\begin{equation}
   [D,f]\psi=i\gamma\cdot\partial f\psi.
\end{equation}
then we need
\begin{equation}
   \lVert [D,f]\rVert=\lVert\sqrt{\partial_\mu f\partial^\mu f}\rVert\le 1.
\end{equation}
This is nothing else than asking the boundedness of the gradient of $f$. We know that $f$ is Lipshitz on the manifold and so, we can apply the Cauchy mean value theorem implying that
\begin{equation}
   \lVert[D,f]\rVert\le\frac{|f(x)-f(y)|}{d_R(x,y)}
\end{equation}
because a constant $L$ exists that can limit the derivatives on the manifold. Now, this implies, due to the condition $\lVert Df\rVert<1$, that
\begin{equation}
   |f(x)-f(y)|\le d_R(x,y)
\end{equation}
and this means that $d_R(x,y)$ is the upper extreme as required by our definition of distance. The main conclusion is that the Dirac operator plays the role of the inverse of the distance $D\sim ds^{-1}$.

\subsection{Quantization of volume}

A noncommutative geometry implies that the volume is quantized with two classes of unity of volume $(1,i)$. This has been recently proved by Connes, Chamseddine and Mukhanov\cite{Chamseddine:2014nxa,Chamseddine:2014uma}. The two classes of volume arise from the fact that the Dirac operator should not be limited to Majorana states in the Hilbert space and so, we need to associate a charge conjugation operator $J$ to our triple $({\cal A},H,D)$. To complete our characterization of our geometry, we recall that the algebra of Dirac matrices implies a $\gamma^5$, the chirality matrix. For an ordinary Riemann manifold, the algebra $\cal A$ is that of functions and is commuting. Remembering that $[D,a]=i\gamma\cdot\partial a$, and noting that, in four dimensions, $x_1,\ x_2,\ x_3,\ x_4$ are legal functions of ${\cal A}$, it is $[D,x_1][D,x_2][D,x_3][D,x_4]=\gamma^1\gamma^2\gamma^3\gamma^4=-i\gamma^5$. For generally chosen functions in ${\cal A}$, $a_0,\ a_1,\ a_2,\ a_3,\ a_4,\ \ldots\ a_d$, summing over all the possible permutations one has a Jacobian, we can define the chirality operator
\begin{equation}
    \gamma=\sum_P(a_0[D,a_1]\ldots[D,a_d]).
\end{equation}
So, in four dimension this gives
\begin{equation}
    \gamma=-iJ\cdot\gamma^5=-i\cdot{\rm det}(e)\gamma^5
\end{equation}
being $J$ the Jacobian, $e^a_\mu$ the vierbein for the Riemann manifold and $\gamma^5=i\gamma^1\gamma^2\gamma^3\gamma^4$ for $d=4$, a well-known result. We used the fact that ${\rm det}(e)=\sqrt{g}$, being $g_{\mu\nu}$ the metric tensor. So, the definition of the chirality operator is proportional to the factor determining the volume of a Riemannian orientable manifold.

In order to see if a Riemannian manifold can be properly quantized, instead of functions we consider operators $Y$ belonging to an operator algebra $\cal A'$. These operators have the properties
\begin{equation}
\label{eq:Y}
   Y^2=\kappa I \qquad Y^\dagger=\kappa Y.
\end{equation}
This is a set of compact operators playing the role of coordinates as in the Heisenberg commutation relations. We have to consider two sets of them $Y_+$ and $Y_-$ as we expect a conjugation of charge operator $C$ to exist such that $CAC^{-1}=Y^\dagger$ for a given operator or complex conjugation for a function. This appears naturally out of a Dirac algebra of gamma matrices. So, a natural way to write down the operators $Y$ is by using an algebra of Dirac matrices $\Gamma^A$ such that
\begin{equation}
   \{\Gamma^A,\Gamma^B\}=2\delta^{AB}, \qquad (\Gamma^A)^*=\kappa\Gamma^A
\end{equation}
with $A,B=1\ldots d+1$, then
\begin{equation}
   Y=\Gamma^AY^A.
\end{equation}
We will have two different set of gamma matrices for $Y_+$ and $Y_-$ that will have independent traces. Using the charge conjugation operator $C$, we can define a new coordinate
\begin{equation}
   Z=2ECEC^{-1}-I
\end{equation}
where $E=(1+Y_+)/2+(1+iY_-)/2$ will project one or the other coordinate. We recognize that the spectrum of $Z$ is in $(1,i)$ given eq.(\ref{eq:Y}). Now, we generalize our equation for the chirality operator imposing a trace on $\Gamma$s both for $Y_+$ and $Y_-$, normalized to the number of components, and we will have
\begin{equation}
\label{eq:Z}
  \frac{1}{n!}\langle Z[D,Z]\ldots[D,Z]\rangle=\gamma.
\end{equation}
where we have introduced the average $\langle\ldots\rangle$ that, in this case, reduces to matrix traces. In order to see the quantization of the volume, let us consider a three dimensional manifold and the sphere $\mathbb{S}^2$. From eq.(\ref{eq:Z}) one has
\begin{equation}
   V_M=\int_M\frac{1}{n!}\langle Z[D,Z]\ldots[D,Z]\rangle d^3x
\end{equation}
and doing the traces one has
\begin{equation}
   V_M=\int_M\left(\frac{1}{2}\epsilon^{\mu\nu}\epsilon_{ABC}Y^A_+\partial_\mu Y^B_+\partial_\nu Y^C_++
	\frac{1}{2}\epsilon^{\mu\nu}\epsilon_{ABC}Y^A_-\partial_\mu Y^B_-\partial_\nu Y^C_-\right)d^3x.
\end{equation}
It is easy to see that this will yield\cite{Chamseddine:2014nxa,Chamseddine:2014uma}
\begin{equation}
   {\rm det}(e^a_\mu)=\frac{1}{2}\epsilon^{\mu\nu}\epsilon_{ABC}Y^A_+\partial_\mu Y^B_+\partial_\nu Y^C_++
	\frac{1}{2}\epsilon^{\mu\nu}\epsilon_{ABC}Y^A_-\partial_\mu Y^B_-\partial_\nu Y^C_-.
\end{equation}
The coordinates $Y_+$ and $Y_-$ belongs to unitary spheres and the Dirac operator has a discrete spectrum, so we are covering all the manifold with a large integer number of these spheres. Thus, the volume is quantized as this condition requires. This can be extended to four dimensions with some more work \cite{Chamseddine:2014nxa,Chamseddine:2014uma}.

\subsection{Stochastic processes on a quantized manifold}

Differently from an ordinary stochastic process, a Wiener process on a quantized manifold will yield the projection of the spectrum $(1,i)$ of the coordinates on the two kind of spheres $Y_+,\ Y_-$. This will depend on the way a particle moves on the manifold taking into account that the distribution of the two kind of unitary volumes is absolutely random. One can construct a process $\Phi$ such that, against a toss of a coin, one gets 1 or i as outcome, assuming the distribution of the unitary volumes is uniform. This can be written
\begin{equation}
   \Phi = \frac{1+B}{2}+i\frac{1-B}{2}
\end{equation}
with $B$ a Bernoulli process producing the value $\pm 1$ depending on the unitary volume hit by the particle such that $B^2=I$, a deterministic process giving always 1, and $\Phi^2=B$. If we want to consider the Brownian motion of the particle on such a manifold we should expect the outcomes to be either $Y_+$ or $Y_-$. So, given the set of $\Gamma$ matrices and the chirality operator $\gamma$, the most general form for a stochastic process on the manifold can be written down (summation on $A$ is implied)
\begin{equation}
\label{eq:ncgSP}
    dY=\Gamma^A\cdot (\kappa_A+\xi_A dX_A\cdot B_A+\zeta_A dt+i\eta_A\gamma^5)\cdot\Phi_A
\end{equation} 
being $\kappa_A,\ \xi_A,\ \zeta_A,\ \eta_A$ arbitrary coefficients of this linear combination. The Bernoulli processes $B_A$ and the Wiener process $dX_A$ cannot be independent. Rather, the sign arising from the Bernoulli process is the same of that of the corresponding Wiener process. This equation provides the equivalent of the eq.(\ref{eq:Y}) for the coordinates on the manifold. This is exactly the formula we will obtain for the fractional powers of a Wiener process. It just represents the motion on a quantized Riemannian manifold with two kind of quanta. Underlying quantum mechanics there appears to be a noncommutative geometry.

\section{Powers of stochastic processes\label{sec2}}

We consider an ordinary Wiener $W$ process describing a Brownian motion and define the $\alpha$-th power of it. We do a proof of existence by construction using a numerical integration technique of a stochastic differential equations (SDE) \cite{fra2}. We will have the process (given $\alpha\in\mathbb{R}^+$) with definition
\begin{equation}
       dX = (dW)^\alpha.
\end{equation}
We build it through the Euler--Maruyama definition of a stochastic process \cite{high} at discrete times
\begin{equation}
\label{eq:e-m}
      X_i=X_{i-1}+(W_i-W_{i-1})^\alpha.
\end{equation}
This is equivalent to take the power and then a (cumulative) sum exactly as done in simulating a Wiener process when $\alpha=1$.

\begin{figure}[!ht]
	\includegraphics{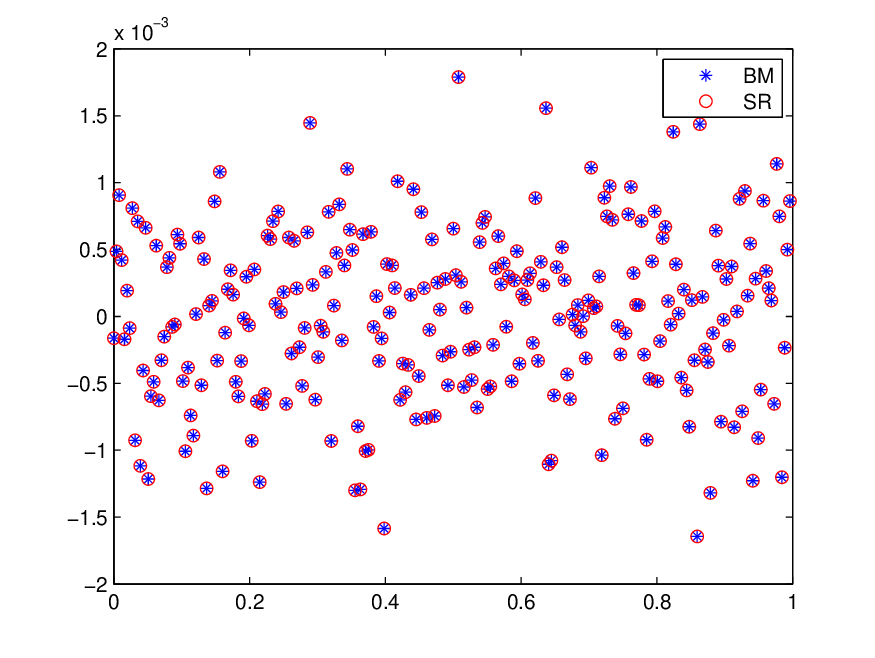}
	\caption{Comparison between the square of the square root process and the original Brownian motion. These coincide perfectly as expected and the square root process exists.
	\label{fig1}}
\end{figure}

For our numerical test we consider the square root process with $\alpha=1/2$ as it is the one of interest for quantum mechanics. In this case the Wiener process has two components: one real and another imaginary. We just compare the original Brownian motion with the square of its square root given by numerically solving eq.(\ref{eq:e-m}). The result is displayed in Fig.~\ref{fig1}.
The results are perfectly identical and our definition by Euler--Maruyama technique just works. The square root process is so shown to exist by construction. We note that the need for a complex valued stochastic process is essential if we aim to recover quantum mechanics. On the other side, taking the square root of values that can have both positive and negative values entails coping with complex numbers. One can always take the power of whatever sequence of numbers as that of a Wiener process.

\section{Square root formula and Fokker--Planck equation\label{sec3}}

Using It\=o calculus to express the square root process with more elementary stochastic processes \cite{okse}, $(dW)^2=dt$, $dW\cdot dt=0$, $(dt)^2=0$ and $(dW)^\alpha=0$ for $\alpha>2$, we could tentatively set
\begin{equation}
\label{eq:sqrt}
    dX=(dW)^\frac{1}{2}\stackrel{?}{=}\left(\mu_0+\frac{1}{2\mu_0}dW\cdot\sgn(dW)-\frac{1}{8\mu_0^3}dt\right)\cdot\Phi_\frac{1}{2}
\end{equation}
being $\mu_0\ne 0$ an arbitrary scale factor and 
\begin{equation}
    \Phi_\frac{1}{2}=\frac{1-i}{2}\sgn(dW)+\frac{1+i}{2}
\end{equation}
a Bernoulli process equivalent to a coin tossing that has the property $(\Phi_\frac{1}{2})^2=\sgn(dW)$. This process is characterized by the values $1$ and $i$ and it is like the Brownian motion went scattering with two different kinds of small pieces of space, each one contributing either 1 or i to the process, randomly. We have introduced the process $\sgn(dW)$ that yields just the signs of the corresponding Wiener process. Eq.(\ref{eq:sqrt}) is unsatisfactory for a reason, taking the square yields
\begin{equation}
   (dX)^2=\mu_0^2\sgn(dW)+dW
\end{equation}
and the original Wiener process is not exactly recovered. We find added a process that has the effect to change the scale of the original Brownian motion while retaining the shape. We can fix this problem by using Pauli matrices. Let us consider two Pauli matrices $\sigma_i,\ \sigma_k$ with $i\ne k$ such that $\{\sigma_i,\sigma_k\}=0$. We can rewrite the above identity as
\begin{equation}
   I\cdot dX=I\cdot(dW)^\frac{1}{2}=\sigma_i\left(\mu_0+\frac{1}{2\mu_0}dW\cdot\sgn(dW)-\frac{1}{8\mu_0^3}dt\right)\cdot\Phi_\frac{1}{2}+i\sigma_k\mu_0\cdot\Phi_\frac{1}{2}
\end{equation}
and so, $(dX)^2=dW$ as it should, after removing the identity matrix on both sides. This idea generalizes easily to higher dimensions using $\gamma$ matrices. In the following we will omit the contribution due to the Pauli matrices but it will be implied to remove the unwanted scale changing process.



Now, let us consider a more general square root process where we assume also a term proportional to $dt$. This forces to take $\mu_0=1/2$ when the square is taken, to recover the original stochastic process, and one has
\begin{equation}
\label{eq:sqrtW}
     dX(t)=[dW(t)+\beta dt]^\frac{1}{2}=
		\left[\frac{1}{2}+dW(t)\cdot\sgn(dW(t))+(-1+\beta\sgn(dW(t)))dt\right]\Phi_{\frac{1}{2}}(t).
\end{equation}
From the Bernoulli process $\Phi_{\frac{1}{2}}(t)$ we can derive
\begin{equation}
     \mu =-\frac{1+i}{2}+\beta\frac{1-i}{2}\qquad\sigma^2=2D=-\frac{i}{2}.
\end{equation}
Then, we get a double Fokker--Planck equation for a free particle, being the distribution function $\hat\psi$ complex valued,
\begin{equation}
\label{eq:fpsch}
     \frac{\partial\hat\psi}{\partial t}=\left(-\frac{1+i}{4}+\beta\frac{1-i}{2}\right)\frac{\partial\hat\psi}{\partial X}
		-\frac{i}{4}\frac{\partial^2\hat\psi}{\partial X^2}.
\end{equation}
This should be expected as we have a complex stochastic process and then two Fokker--Planck equations are needed to describe it. We have obtained an equation strongly resembling the Schr\"odinger equation for a complex distribution function. We can ask at this point if indeed are recovering quantum mechanics. In the following section we will perform a numerical check of this hypothesis.

\section{Recovering the kernel of the Schr\"odinger equation\label{sec4}} 

\begin{figure}[!ht]
	\includegraphics{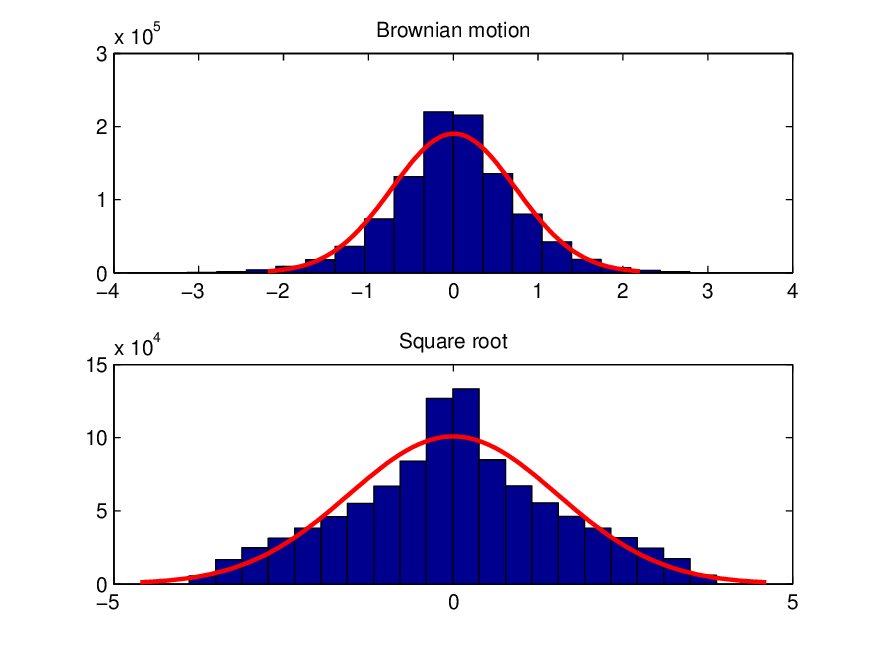}
	\caption{Comparison between the distributions of the Brownian motion and its square root after a Wick rotation.}
	\label{fig2}
\end{figure}

If really the square root process diffuses as a solution of the Schr\"odinger equation we should be able to recover the corresponding solution for the kernel
\begin{equation} 
    \hat\psi=(4\pi it)^{-\frac{1}{2}}\exp{\left(ix^2/4t\right)}
\end{equation}
sampling the square root process. To see this we note that a Wick rotation, $t\rightarrow -it$, turns it into a heat kernel as we get immediately
\begin{equation}  
     K=(4\pi t)^{-\frac{1}{2}}\exp{\left(-x^2/4t\right)}.
\end{equation}
A Montecarlo simulation can be easily executed extracting the square root of a Brownian motion and, after a Wick rotation, to show that a heat kernel is obtained. We have generated 10000 paths of Brownian motion and extracted its square root in the way devised in Sec.~\ref{sec2}. We have evaluated the corresponding distribution after Wick rotating the results for the square root. The Wick rotation generates real results as it should be expected and a comparison can be performed. The result is given in Fig.~\ref{fig2}
The quality of the fit can be evaluated being $\hat\mu=0.007347$ with confidence interval $[{0.005916},{0.008778}]$, $\hat\sigma=0.730221$ with confidence interval $[{0.729210},{0.731234}]$ for the heat kernel while one has $\hat\mu=0.000178$ with confidence interval $[{-0.002833},{0.003189}]$ and $\hat\sigma=1.536228$ with confidence interval $[{1.534102},{1.538360}]$ for the Schr\"odinger kernel. Both are centered around 0 and there is a factor $\sim 2$ between standard deviations as expected from eq.~(\ref{eq:fpsch}). Both the fits are exceedingly good. Having recovered the Schr\"odinger kernel from Brownian motion with the proper scaling factors in mean and standard deviation, we can conclude that we are doing quantum mechanics: {\sl Square root of a Brownian motion describes the motion of a quantum particle}. Need for Pauli matrices, as shown in the preceding section, implies that spin cannot be neglected.

\section{Particle in a potential\label{sec5}}

In order to understand how to introduce a potential within this approach we use the following mapping theorem between the Fokker--Planck and the Schr\"odinger equation \cite{pavl,risk}:

\begin{theo}
The Fokker--Planck operator for a gradient flow can be written in the self-adjoint form
\begin{equation}
   \frac{\partial\hat\psi}{\partial t}=D\nabla\cdot\left(e^{-\frac{U}{D}}\nabla\left(e^{\frac{U}{D}}\hat\psi\right)\right).
\end{equation}
{\it Define now $\psi(x,t)=e^{\frac{U}{2D}}\hat\psi(x,t)$. Then $\psi$ solves the PDE}
\begin{equation}
   \frac{\partial\psi}{\partial t}=D\Delta_2\psi-V(x)\psi,\qquad V(x):=\frac{|\nabla U|^2}{4D}-\frac{\Delta_2 U}{2}.
\end{equation}
\end{theo}

On the basis of the given theorem, we can immediately generalize our formulation to the case of a potential. We will have
\begin{equation}
     dX(t)=[dW(t)+U(X,t)dt]^\frac{1}{2}=\left\{\frac{1}{2}+dW\cdot\sgn(dW(t))+
    (-1+U(X,t)\sgn(dW(t)))dt\right\}\Phi_{\frac{1}{2}}(t).
\end{equation}
The corresponding Fokker--Planck equation will be
\begin{equation}
   \frac{\partial\hat\psi}{\partial t}=\frac{\partial}{\partial X}\left[\left(-\frac{1+i}{4}
   +\frac{1-i}{4}U(X,t)\right)\hat\psi\right]-\frac{i}{4}\frac{\partial^2\hat\psi}{\partial X^2}.
\end{equation}
As an example we consider a harmonic oscillator with $U(X)=kX^2/2$
\begin{equation}
     dX(t)=\left[dW(t)+\frac{k}{2}X^2dt\right]^\frac{1}{2}=\left\{\frac{1}{2}+dW\cdot\sgn(dW(t))+
    \left(-1+\frac{k}{2}X^2\sgn(dW(t))\right)dt\right\}\Phi_{\frac{1}{2}}(t).
\end{equation}
Here $k$ is an arbitrary constant and the quantum potential is $V(X)=|k|^2X^2-\frac{k}{2}$, using the mapping between the Fokker--Planck and the Schr\"odinger equations.
The corresponding Schr\"odinger equation will be
\begin{equation}
-i\frac{\partial\psi^*}{\partial t}=-\frac{1}{4}\frac{\partial^2\psi^*}{\partial X^2}
	+\left(|k|^2X^2-\frac{k}{2}\right)\psi^*
\end{equation}
with the introduction of $\psi^*$ as we get what is conventionally a time-reversed quantum evolution.

\section{Square root and noncommutative geometry\label{sec6}}


We have seen that, in order to extract the square root of a stochastic process, we needed Pauli matrices or, generally speaking, a Clifford algebra. This idea was initially put forward by Dirac to derive his relativistic equation for fermions.
The simplest and non-trivial choice is obtained, as said above, using Pauli matrices $\{\sigma_k\in C\ell_3(\mathbb{C}),\ k=1,2,3\}$ that satisfy
\begin{equation}
    \sigma_i^2=I \qquad \sigma_i\sigma_k=-\sigma_k\sigma_i\qquad i\ne k.
\end{equation}
This proves to be insufficient to go to dimensions higher than 1+1 for Brownian motion. The more general solution is provided by a Dirac algebra of $\gamma$ matrices $\{\gamma_k\in C\ell_{1,3}(\mathbb{C}),\ k=0,1,2,3\}$ such that
\begin{equation}
    \gamma_0^2=I \qquad \gamma_1^2=\gamma_2^2=\gamma_3^2=-I\qquad \gamma_i\gamma_k+\gamma_k\gamma_i=2\eta_{ik}
\end{equation}
being $\eta_{ik}$ the Minkowski metric. In this way one can introduce three different Brownian motions for each spatial coordinates and three different Bernoulli processes for each of them. The definition is now
\begin{equation}
   dE=\sum_{k=1}^3i\gamma_k\left(\mu_k+\frac{1}{2\mu_k}|dW_k|-\frac{1}{8\mu_k^3}dt\right)\cdot\Phi_\frac{1}{2}^{(k)}
	+\sum_{k=1}^3i\gamma_0\gamma_k\mu_k\Phi_\frac{1}{2}^{(k)}
\end{equation}
It is now easy to check that
\begin{equation}
    (dE)^2=I\cdot(dW_1+dW_2+dW_3).
\end{equation}
The Fokker-Planck equations have a solution with 4 components, as now the distribution functions are Dirac spinors. These are given by
\begin{equation}
     \frac{\partial\hat\Psi}{\partial t}=\sum_{k=1}^3\frac{\partial}{\partial X_k}\left(\mu_k\hat\Psi\right)
		-\frac{i}{4}\Delta_2\hat\Psi
\end{equation}
being $\mu_k=-\frac{1+i}{4}+\beta_k\frac{1-i}{2}$. This implies that, the general formula for the square root process implies immediately spin and antimatter for quantum mechanics that now come out naturally. But this appears just like the non-relativistic limit of the Dirac equation and so, having already introduced the $\gamma$ matrices at this stage, it should be natural to get a fully covariant Dirac equation. In the next section we will show that this indeed the case so that, the metric element of a noncommutative geometry arise naturally as the Fokker--Planck equation of a stochastic process.

\section{Dirac equation\label{sec7}}

Dirac equation works on a 4-dimensional manifold and so, we will need four Wiener processes to derive it. This assures full Lorentz invariance but, on the other side, time should be treated as any other space variable. We need a further time variable, a fictitious one (as happens in stochastic quantization), to get Fokker--Planck equations in this case. To accomplish this one has to introduce the $\gamma_5$ matrix, as already seen in noncommutative geometry, in the following way
\begin{equation}
\label{eq:Dirac}
     dE=\sum_{k=0}^3i\gamma^k\left(\mu_k+\frac{1}{2\mu_k}|dW_k|-\frac{1}{8\mu_k^3}dt\right)\cdot\Phi_\frac{1}{2}^{(k)}+\sum_{k=0}^3i\gamma^5\gamma^k\mu_k\Phi_\frac{1}{2}^{(k)}.
\end{equation}
Now one has a fictitious time variable $\tau$ but we have a full family of solutions to the Fokker-Planck equations parametrized by $\tau$. Only the fixed point solutions, the eigenstate with zero eigenvalue, reproduce the Klein-Gordon equation for a free massless particle with a Dirac spinor.
\begin{equation}
     \frac{\partial\hat\Psi}{\partial\tau}=\partial\cdot\left(\mu\hat\Psi\right)-\frac{i}{4}\partial^2\hat\Psi.
\end{equation}
This recovers completely Dirac theory for a free particle from Brownian motions. We recognize in eq.(\ref{eq:Dirac}) the same stochastic process arising in noncommutative geometry in eq.(\ref{eq:ncgSP}).

\section{Conclusions\label{sec8}}

We have shown the existence of a class of stochastic processes that can support quantum behavior. A typical one is the square root of a Brownian motion from which the Schr\"odinger equation comes out naturally. The case with a potential was also discussed and applied to the harmonic oscillator. Finally, we have derived the Dirac equation while spin and antimatter are naturally introduced by a stochastic behavior. This formalism could entail a new understanding of quantum mechanics and give serious hints on the properties of space-time for quantum gravity. This yields a deep connection with noncommutative geometry as formulated by Alain Connes through the more recent proposal of space quantization by Connes himself, Chamseddine and Mukhanov. This quantization of volume entails two kind of quanta implying naturally the unity $(1,i)$ that arises in the square root of a Wiener process. Indeed, a general stochastic process for a particle moving on such a quantized volume corresponds to our formula of the square root of a stochastic process on a 4-dimensional manifold. Spin appears to be an essential ingredient, already at a formal level, to treat such fractional powers of Brownian motion.

Finally, it should be interesting, and rather straightforward, to generalize this approach to a Dirac equation on a generic manifold. The idea would be to recover also Einstein equations as a fixed point solution to the Fokker-Planck equations as already happens in string theory. Then they would appear as a the result of a thermodynamic system at the equilibrium based on noncommutative geometry. This is left for further study. 

\begin{acknowledgments}
I would like to thank Alfonso Farina for giving me the chance to unveil some original points of view on this dusty corner of quantum physics.
\end{acknowledgments}

\end{document}